\documentclass[aps,prd,onecolumn,amsmath,amssymb,superscriptaddress,floatfix,nofootinbib]{revtex4-2}

\usepackage{graphicx}
\usepackage{subcaption}
\usepackage{multirow}
\usepackage{booktabs}
\usepackage{rotating}
\usepackage{tabularx}
\usepackage{bm}

\begin{document}

\title{Mass spectrum, magnetic moments and Regge trajectories of $\Omega_{ccb}$ and $\Omega_{cbb}$ baryons in the nonrelativistic quark--diquark model}

\author{\"{O}znur \c{C}ak{\i}r}
\email{znurcaa@gmail.com}
\affiliation{Department of Physics, Faculty of Science, Ondokuz May{\i}s University, 55139 Atakum, Samsun, T\"{u}rkiye}
\affiliation{\c{S}ehit \"{O}zg\"{u}r Gencer Girls Anatolian Imam Hatip High School, 58400 \c{S}ark{\i}\c{s}la, Sivas, T\"{u}rkiye}

\author{Halil Mutuk}
\email{hmutuk@omu.edu.tr}
\affiliation{Department of Physics, Faculty of Science, Ondokuz May{\i}s University, 55139 Atakum, Samsun, T\"{u}rkiye}

\begin{abstract}
In this work, we investigate the mass spectra, magnetic moments, and Regge trajectories of the triply heavy baryons $\Omega_{ccb}$ and $\Omega_{cbb}$ within a nonrelativistic constituent quark model based on the quark--diquark approximation, which reduces the three-body problem to an effective two-body system. For each baryon, all three possible diquark clusterings are considered, providing a qualitative indication of the sensitivity of the results to the quark--diquark decomposition. The model parameters are fixed by a fit to the measured $B_c$ meson spectrum, thereby anchoring the baryon predictions to experimentally constrained inputs and establishing a consistent link between the heavy meson and baryon sectors. We obtain ground-state masses of approximately $8.0$~GeV for $\Omega_{ccb}$ and $11.0$~GeV for $\Omega_{cbb}$, with radial and orbital excitation patterns in good agreement with the results reported in the literature. The computed magnetic moments of the spin-$\tfrac{1}{2}$ and spin-$\tfrac{3}{2}$ states are consistent with the results of various approaches. A radial Regge analysis in the $(n_r, M^2)$ plane reveals approximately linear $P$-wave trajectories and mildly curved $S$-wave trajectories, with slope and intercept parameters that scale systematically with the heavy-quark content of the baryon. These results suggest that the nonrelativistic quark--diquark framework provides a reliable description of triply heavy baryons and serves as a useful reference for future experimental searches, particularly at LHCb.
\end{abstract}

%\keywords{triply heavy baryons, $\Omega_{ccb}$, $\Omega_{bbc}$, nonrelativistic potential, quark--diquark, mass spectrum, magnetic moment}

\maketitle

\section{Introduction}
\label{sec:intro}

The study of hadrons containing heavy quarks provides a unique window into the dynamics of Quantum Chromodynamics (QCD), the fundamental theory of the strong interaction. Among these, triply heavy baryons---bound states of three charm $(c)$ or bottom $(b)$ quarks---occupy a special place due to their distinctive properties. Unlike conventional baryons with light or mixed heavy-light quark content, triply heavy baryons are predominantly governed by perturbative QCD effects. The large masses of the heavy quarks suppress relativistic corrections, making these systems particularly amenable to theoretical treatment while still retaining essential nonperturbative features such as confinement and spin-dependent interactions.

From a theoretical standpoint, triply heavy baryons serve as an ideal testing ground for QCD in the heavy-quark limit. Their spectroscopy can be compared with lattice QCD simulations, which are particularly precise in this regime due to reduced systematic uncertainties associated with quark discretization. Additionally, the nonrelativistic nature of these systems allows for the application of potential models, heavy-quark effective theory (HQET), and even semi-analytical approaches to estimate their properties. Understanding their structure also sheds light on more general aspects of QCD, such as the nature of quark-quark interactions in the absence of light quarks and the role of three-body forces in baryonic systems.

Unlike their mesonic counterparts (e.g., charmonium or bottomonium), triply heavy baryons offer complementary insights into the quark-quark interaction dynamics, free from the complications of light-quark sea effects. Despite their theoretical importance, experimental observation of triply heavy baryons remains elusive due to their suppressed production cross-sections in high-energy collisions. Nevertheless, their predicted properties---ground-state masses, excitations, and decay patterns---are critical for guiding future searches at facilities like the LHC, Belle II, and BES III.

Triply heavy baryons are studied by using various methods such as quark models~\cite{Silvestre-Brac:1996myf,Vijande:2004at,Migura:2006ep,Jia:2006gw,Martynenko:2007je,Roberts:2007ni,Patel:2008mv,Flynn:2011gf,Shah:2017jkr,Weng:2018mmf,Qin:2018dqp,Yang:2019lsg,Liu:2019vtx,Faustov:2021qqf,Ortiz-Pacheco:2023kjn,Yu:2025gdg,Zhou:2025fpp,Mohan:2026rcg}, lattice QCD~\cite{Dhindsa:2024erk,Padmanath:2013zfa,Mathur:2018epb,Brown:2014ena,Burch:2015pka,Meinel:2010pw,Meinel:2012qz,Alexandrou:2023dlu,Li:2022vbc,Lyu:2021qsh,Bahtiyar:2020uuj,Alexandrou:2017xwd,Chen:2017kxr,Can:2015exa,Alexandrou:2014sha,PACS-CS:2013vie,Durr:2012dw,Alexandrou:2012xk,Briceno:2012wt,Chiu:2005zc}, potential nonrelativistic QCD (pNRQCD)~\cite{Brambilla:2005yk,Llanes-Estrada:2011gwu}, QCD sum rules (QCDSR)~\cite{Zhang:2009re,Wang:2011ae,Wang:2020avt,Najjar:2025dzl}, Regge phenomenology~\cite{Guo:2008he,Wei:2015gsa,Wei:2016jyk} and some other methods~\cite{Serafin:2018aih,Yin:2019bxe,Gutierrez-Guerrero:2019uwa,Gomez-Rocha:2023jfr,Celiberto:2025ogy,Celiberto:2025csa}. These studies contribute to the understanding of triply heavy baryons and hadron physics.

The constituent quark model (CQM) has proven to be an effective framework for describing hadron spectroscopy, successfully predicting masses, decay properties, and spin structures of mesons and baryons across different flavor sectors. In this approach, quarks are treated as quasiparticles with effective masses generated dynamically through spontaneous chiral symmetry breaking, interacting via phenomenological potentials that incorporate key QCD-inspired mechanisms, including one-gluon exchange (OGE) and a confining potential. For triply heavy baryons, the CQM offers a particularly clean theoretical setup since the heavy quark masses minimize complications arising from relativistic motion and coupled-channel effects that are significant in light-quark systems.

Triply heavy baryons also constitute an important testing ground for symmetry principles and conservation laws~\cite{Cheng:2021qpd}. Their fundamental properties, including masses and magnetic moments, offer valuable insight into hadronic structure and the validity of different theoretical approaches~\cite{ExHIC:2017smd}. Moreover, these systems provide a simplified yet nontrivial framework for investigating the three-body problem, where the quark--diquark approximation reduces it to an effective two-body system while preserving key physical features~\cite{Faustov:2021qqf,Roberts:2007ni,Yu:2025gdg}.

In this work, the mass spectra and magnetic moments of $\Omega_{ccb}$ and $\Omega_{cbb}$ are determined using parameters calibrated to the spectroscopy of $B_c$ mesons. Within a nonrelativistic quark model, the three-body problem is treated in the quark--diquark approximation, providing a consistent framework for heavy-quark systems. The calibration to the $B_c$ spectrum constrains the parameter space and improves the internal consistency of the model. By solving the Schr\"{o}dinger equation with these fitted parameters, we obtain systematic predictions for both ground and excited states, as well as for the magnetic moments of spin-$\tfrac{1}{2}$ and spin-$\tfrac{3}{2}$ baryons. The results are compared with existing theoretical studies. Regge trajectories are constructed in both the $(J,M^2)$ and $(n,M^2)$ planes to analyze the systematic behavior of triply heavy baryons.

The remainder of this paper is organized as follows. In Sec.~\ref{sec:theoretical_framework}, we present the theoretical framework. Section~\ref{sec:numresdis} is devoted to the numerical results and their discussion. Finally, Sec.~\ref{sec:conclusion} summarizes the main findings and outlines directions for future work.

\section{Theoretical Framework}
\label{sec:theoretical_framework}

\subsection{Diquarks as constituents of triply heavy baryons}
\label{subsec:diquark_motivation}

A baryon composed of three quarks $Q_1 Q_2 Q_3$ constitutes a quantum-mechanical three-body problem. A considerable simplification is achieved if two of the quarks form a tightly bound diquark, reducing the baryon to an effective two-body system.

The basis of the diquark concept is the color decomposition $\mathbf{3}\otimes\mathbf{3}=\bar{\mathbf{3}}\oplus\mathbf{6}$. The antitriplet channel is attractive, with a color factor $\kappa=-2/3$ that is half the quark--antiquark value $\kappa=-4/3$, while the sextet is repulsive~\cite{Debastiani:2017msn}. Combining an antitriplet diquark with a triplet quark, $\mathbf{3}\otimes\bar{\mathbf{3}}=\mathbf{1}\oplus\mathbf{8}$, yields a color-singlet baryon. The diquark is thus not an asymptotic state of QCD but an effective degree of freedom encoding the dominant two-body correlations within the baryon, consistent with gauge symmetry~\cite{Anselmino:1992vg,Jaffe:2004ph,Wilczek:2004im}.

Ground-state ($L=0$) diquarks come in two types: scalar $[QQ']$ ($^1S_0$, $J^P=0^+$, spin-antisymmetric) and axial-vector $\{QQ'\}$ ($^3S_1$, $J^P=1^+$, spin-symmetric). In the classification of Jaffe~\cite{Jaffe:2004ph}, these are called ``good'' and ``bad'' diquarks, respectively, reflecting the stronger chromomagnetic attraction in the scalar channel. For the equal-flavor diquarks $cc$ and $bb$, the Pauli principle---combined with the antisymmetric color-antitriplet and symmetric spatial wave functions---requires a symmetric spin wave function, selecting only the axial-vector configurations $\{cc\}$ and $\{bb\}$; the scalar configurations $[cc]$ and $[bb]$ are forbidden~\cite{Jaffe:2004ph,Selem:2006nd}. For the mixed-flavor diquark $bc$, the two quarks are distinguishable and both $[bc]$ ($S_d=0$) and $\{bc\}$ ($S_d=1$) are allowed. The chromomagnetic hyperfine interaction splits the $[bc]$ and $\{bc\}$ masses, generating additional states in the baryon spectrum. We consider both configurations in this work.

The quark-diquark decompositions for the baryons studied here are:
\begin{itemize}
    \item $\Omega_{ccb}$: $\;b + \{cc\}$, $\;c + [bc]$, $\;c + \{bc\}$,
    \item $\Omega_{cbb}$: $\;c + \{bb\}$, $\;b + [bc]$, $\;b + \{bc\}$.
\end{itemize}
Since the two-body reduction explicitly breaks the permutation symmetry of the three-body problem, different clusterings of the same baryon generally yield different mass predictions. The spread among them provides a quantitative estimate of the systematic uncertainty inherent to the quark-diquark approximation, and the availability of multiple channels for both $\Omega_{ccb}$ and $\Omega_{cbb}$ thus offers internal consistency checks that are not possible for equal-flavor systems.

Several limitations of the model should be acknowledged. First, only the spin-spin component of the Breit--Fermi Hamiltonian is retained; the spin-orbit ($\mathbf{L}\cdot\mathbf{S}$) and tensor interactions are neglected. These terms vanish for $L=0$ ground states, so their omission does not affect ground-state predictions, but for $L\geq 1$ excitations the predicted masses should be understood as spin-averaged values with unresolved fine structure. Second, relativistic recoil corrections, including the Thomas precession, can become non-negligible for asymmetric systems such as $c+\{bb\}$ or $b+\{cc\}$; their systematic inclusion would require a semi-relativistic treatment. Third, corrections from quark exchange, diquark excitation, and finite-size effects, which can be significant for systems with light quarks~\cite{Silvestre-Brac:1996myf,Valcarce:2008dr}, are expected to be suppressed here by the large quark masses~\cite{Ebert:2002ig,Roberts:2007ni}. This expectation is supported by lattice QCD calculations of triply heavy baryon masses~\cite{Brown:2014ena,Mathur:2018epb}, which solve the full three-body problem from first principles; the agreement between such results and quark-diquark predictions reported in the literature~\cite{Ebert:2002ig,Roberts:2007ni} suggests that the dominant correlations are well captured by the diquark approximation.

A practical advantage of the framework is the direct connection it establishes between the meson and baryon sectors. The same nonrelativistic potential, with modifications dictated solely by color algebra, is applied first to the quark--quark channel to obtain $M_{\{cc\}}$, $M_{\{bb\}}$, $M_{[bc]}$, and $M_{\{bc\}}$, and then to the quark-diquark channel to predict the baryon masses. This ensures that all predictions for $\Omega_{ccb}$ and $\Omega_{cbb}$ are anchored to the experimentally constrained meson spectrum with no additional free parameters.

The diquark masses were obtained by numerically solving the Schr\"odinger equation for the heavy--heavy quark subsystem within the same Cornell-type potential framework employed in the meson sector. In order to extend the quark--antiquark interaction to the quark--quark (diquark) system, the change in the SU(3) color structure was taken into account. Since the attractive color antitriplet diquark configuration corresponds to a color factor reduced by one-half compared to the color-singlet quark--antiquark system, the substitutions
$\kappa \rightarrow \frac{\kappa}{2}, \qquad b \rightarrow \frac{b}{2}$
were applied in the Cornell potential. Accordingly, the diquark mass was calculated as the sum of the constituent quark masses and the corresponding energy eigenvalue of the two-quark system,
\begin{equation}
M_d = m_{Q_1}+m_{Q_2}+E_{QQ},
\end{equation}
where \(E_{QQ}\) denotes the energy eigenvalue of the heavy diquark subsystem. The same set of parameters determined from the \(B_c\) meson spectrum was used
without introducing additional free parameters in the diquark sector.

\subsection{Potential Model}
\label{subsec:potential_model}

We now consider the interaction between a heavy quark $Q$ and a heavy diquark $d\equiv[Q'Q'']_{\bar{\mathbf{3}}}$, where $Q,Q',Q''=c,b$. The color coupling of a fundamental-triplet quark and an antitriplet diquark decomposes as $\mathbf{3}\otimes\bar{\mathbf{3}}=\mathbf{1}\oplus\mathbf{8}$, so that the color-singlet channel carries the same color factor $\kappa=-4/3$ as for the quark--antiquark system. Consequently, the unperturbed quark-diquark potential retains the Cornell form~\cite{Eichten:1979ms},
\begin{equation}
    V(r) = \frac{\kappa\alpha_S}{r} + br,
    \label{eq:CornellQd}
\end{equation}
with the same values of $\alpha_S$ and $b$ as determined from the meson fits, and with the Coulomb term $V_V(r)\equiv\kappa\alpha_S/r$ arising from the effective OGE between the quark and the diquark. The linear term accounts for the confinement of the quark-diquark system. A nonrelativistic treatment is well justified here, since both the quark and the diquark are heavy, ensuring that the kinetic energy remains much smaller than the rest masses of the constituents.

We formulate the Schr\"{o}dinger equation in the center-of-mass frame of the quark-diquark system. Using spherical coordinates, the angular and radial parts can be factorized. Let $\mu_{Qd}\equiv m_Q M_d/(m_Q+M_d)$, where $m_Q$ is the constituent mass of the quark and $M_d$ is the mass of the diquark as determined in the previous stage. The time-independent radial Schr\"{o}dinger equation then reads
\begin{equation}
    \left\{  -\frac{1}{2\mu_{Qd}}\left[\frac{d^2}{dr^2}+\frac{2}{r}\frac{d}{dr}-\frac{L(L+1)}{r^2}\right]+V(r) \right\} \psi(r) = E_{Qd}\,\psi(r),
    \label{eq:SchrodingerQd}
\end{equation}
with the orbital quantum number $L$ and the quark-diquark energy eigenvalue $E_{Qd}$. Substituting $\psi(r) \equiv r^{-1}\varphi(r)$, Eq.~(\ref{eq:SchrodingerQd}) transforms into
\begin{equation}
    \left\{  \frac{1}{2\mu_{Qd}}\left[-\frac{d^2}{dr^2}+\frac{L(L+1)}{r^2}\right] +V(r) \right\} \varphi(r) = E_{Qd}\,\varphi(r).
    \label{eq:SchrodingerRedQd}
\end{equation}

In the quark-diquark picture, the diquark is treated as an effective point-like color-antitriplet object carrying spin $\mathbf{S}_d$. A spin-spin interaction between the quark spin $\mathbf{S}_Q$ and the diquark spin $\mathbf{S}_d$ is then included in analogy with the quark--antiquark case. Based on the Breit--Fermi Hamiltonian for OGE~\cite{Lucha:1991vn,Lucha:1995zv,Voloshin:2007dx,Eiglsperger:2007ay,Debastiani:2017msn}, this interaction takes the form
\begin{equation}
    V_S(r) = -\frac{2}{3(2\mu_{Qd})^2}\nabla^2V_V(r)\, \langle\mathbf{S}_Q\cdot\mathbf{S}_d \rangle
    = -\frac{2\pi\kappa\alpha_S}{3\mu_{Qd}^2}\delta^3(r)\, \langle\mathbf{S}_Q\cdot\mathbf{S}_d \rangle.
    \label{eq:spinQd1}
\end{equation}

As in the meson sector, we replace the Dirac delta function with a smeared Gaussian function regulated by the parameter $\sigma$~\cite{Godfrey:1985xj},
\begin{equation}
    V_S(r) = -\frac{2\pi\kappa\alpha_S}{3\mu_{Qd}^2}\left(\frac{\sigma}{\sqrt{\pi}}\right)^3\exp\left(-\sigma^2 r^2\right) \langle\mathbf{S}_Q\cdot\mathbf{S}_d \rangle,
    \label{eq:spinQd2}
\end{equation}
where the value of $\sigma$ is inherited from the meson fits. Incorporating this spin-spin interaction, Eq.~(\ref{eq:SchrodingerRedQd}) takes the form
\begin{equation}
    \left[  -\frac{d^2}{dr^2} +V_{\mathrm{eff}}(r) \right]\varphi(r) = 2\mu_{Qd}\, E_{Qd}\,\varphi(r),
    \label{eq:SchrodingerRedQd2}
\end{equation}
where the effective potential $V_{\mathrm{eff}}(r)$ is given by
\begin{equation}
    V_{\mathrm{eff}}(r)\equiv 2\mu_{Qd}\left[ V(r)+V_S(r) \right] + \frac{L(L+1)}{r^2},
    \label{eq:VeffQd}
\end{equation}
taking into account the quark-diquark spin-spin interaction. For axial-vector diquarks with spin $S_d=1$ and a heavy quark with spin $S_Q=1/2$, the product $\mathbf{S}_Q\cdot\mathbf{S}_d$ is evaluated in the eigenbasis of the total spin $\mathbf{S}=\mathbf{S}_Q+\mathbf{S}_d$ as
\begin{equation}
    \mathbf{S}_Q\cdot\mathbf{S}_d = \frac{1}{2}\left[S(S+1)-S_Q(S_Q+1)-S_d(S_d+1)\right],
    \label{eq:SdotSQd}
\end{equation}
yielding $\mathbf{S}_Q\cdot\mathbf{S}_d = 1/4$ for $S=3/2$ and $\mathbf{S}_Q\cdot\mathbf{S}_d = -3/4$ for $S=1/2$. Equation~(\ref{eq:SchrodingerRedQd2}) is solved numerically for the energy eigenvalue $E_{Qd}$ and the reduced wave function $\varphi(r)$, using the same Dirichlet boundary conditions at $r=0$ and $r=r_0$ as in the meson and diquark sectors. The mass $M_B$ of the triply heavy baryon is then obtained as
\begin{equation}
    M_B = m_Q + M_d + E_{Qd}.
    \label{eq:baryonmass}
\end{equation}

\subsection{Model parameters}
\label{subsec:numerical_method}

The free parameters of the model, $\boldsymbol{\nu}\equiv(m_c,\,m_b,\,\alpha_S,\,b,\,\sigma)$, are determined by fitting to the known mass spectrum of $B_c$ mesons~\cite{ParticleDataGroup:2024cfk}. Since $B_c$ states are composed of a charm and a bottom quark, they probe the same heavy-quark potential that governs the diquark and quark-diquark dynamics studied in this work, making them a natural calibration system. This choice also provides a consistent bridge between the meson and baryon sectors and partially compensates for the absence of experimental data on triply heavy baryons.

The fit is performed by minimizing the $\chi^2$ function
\begin{equation}
    \chi^2(\boldsymbol{\nu}) = \sum_{i=1}^{N} \left[ M_i^{\mathrm{exp}} - M_i^{\mathrm{th}}(\boldsymbol{\nu}) \right]^2,
    \label{eq:chi2}
\end{equation}
where the sum runs over the $N$ experimentally measured $B_c$ meson masses $M_i^{\mathrm{exp}}$ and $M_i^{\mathrm{th}}(\boldsymbol{\nu})$ denotes the corresponding model prediction for a given parameter set $\boldsymbol{\nu}$. The minimization proceeds in two stages. First, a coarse scan of the parameter space is carried out by evaluating $\chi^2$ on a dense random grid to identify the region containing the global minimum. Second, an iterative adaptive refinement is performed within this region, progressively narrowing the search domain until the best-fit point is localized.

For each trial parameter set, the reduced radial Schr\"{o}dinger equation~(\ref{eq:SchrodingerRedQd2}) is solved numerically subject to Dirichlet boundary conditions at $r=0$ and at a sufficiently large cutoff radius $r_0$, chosen such that the energy eigenvalue $E$ is stable to five significant figures. The theoretical meson mass entering Eq.~(\ref{eq:chi2}) is then obtained as $M_i^{\mathrm{th}} = m_1 + m_2 + E_i$, where $m_1$ and $m_2$ are the constituent quark masses and $E_i$ is the energy eigenvalue of the corresponding state. The same numerical procedure is subsequently applied in the diquark and quark-diquark sectors, with the baryon mass given by
\begin{equation}
    M_B = m_Q + M_d + E_{Qd},
    \label{eq:mass_num}
\end{equation}
where $m_Q$ is the spectator quark mass, $M_d$ the diquark mass, and $E_{Qd}$ the quark-diquark energy eigenvalue.

The resulting parameters are summarized in Table~\ref{tab:parameters}. These values yield a satisfactory reproduction of the experimental $B_c$ spectrum and are used without further adjustment in all subsequent diquark and baryon calculations.

\begin{table}[h]
\centering
\caption{Model parameters used in this work.}
\label{tab:parameters}
\begin{tabular}{lc}
\hline\hline
Parameter & Value \\
\hline
$m_c$       & $1.5366$~GeV \\
$m_b$       & $4.6047$~GeV \\
$\alpha_s$  & $0.4268$ \\
$b$         & $0.1840$~GeV$^2$ \\
$\sigma$    & $0.2614$~GeV \\
\hline\hline
\end{tabular}
\end{table}

Using this set of parameters, we calculated the masses of the observed $B_c$ meson states. The results are presented in Table~\ref{tab:bccomparison}.

\begin{table}[h]
\centering
\caption{Comparison of calculated $B_c$ meson states with experimental ones. The results are presented in MeV.}
\label{tab:bccomparison}
\begin{tabular}{lcc}
\hline\hline
State & This work & Experimental~\cite{ParticleDataGroup:2024cfk} \\
\hline
$B_c(1S)$ & $6271$ & $6274.47 \pm 0.27 \pm 0.17$ \\
$B_c(2S)$ & $6871$ & $6871.2 \pm 1.0$ \\
\hline\hline
\end{tabular}
\end{table}

As seen in Table~\ref{tab:bccomparison}, the model reproduces the two experimentally known $B_c$ states with remarkable accuracy. For the ground state $B_c(1S)$, the predicted mass of $6271$~MeV deviates from the measured value of $M=6274.47\pm 0.27\pm 0.17$~MeV by approximately $3.5$~MeV, corresponding to a relative discrepancy of less than $0.06\%$. For the radially excited state $B_c(2S)$, the agreement is even closer: the calculated mass of $6871$~MeV differs from the experimental value of $M=6871.2 \pm 1.0$~MeV by only $0.2$~MeV, well within the experimental uncertainty. The $1S$--$2S$ mass splitting, which is sensitive to the interplay between the Coulomb and confining terms in the Cornell potential, is predicted to be $600$~MeV, in excellent agreement with the experimental value of $(596.7\pm 1.0)$~MeV.

\subsection{Magnetic Moments}
\label{subsec:magnetic_moments}

The magnetic moment is a fundamental static property of a hadron that encodes information about the distribution of charge and spin among its constituents. For baryons, it provides a direct probe of the internal quark dynamics: the relative orientation and weighting of individual quark magnetic moments within the baryon wave function reflect the underlying spin--flavor structure in a way that is complementary to the mass spectrum. Historically, the measurement of nucleon magnetic moments played a pivotal role in establishing the quark model~\cite{Gell-Mann:1964ewy}, and their successful description remains a benchmark for any model of hadron structure~\cite{Beg:1964nm,DeRujula:1975qlm}.

For triply heavy baryons, magnetic moments carry particular theoretical interest for several reasons. First, since all constituents are heavy, the nonrelativistic expansion is well controlled and the leading-order constituent quark model prediction is expected to be reliable, with relativistic corrections suppressed by powers of $v/c$. Second, the magnetic moment is sensitive to the effective quark masses inside the baryon, which differ from the free constituent masses due to binding effects, thereby providing an independent constraint on the interquark dynamics beyond the mass spectrum alone. Third, the ratios and signs of the magnetic moments across the $J=1/2$ and $J=3/2$ multiplets directly test the spin structure of the baryon wave function and, in the quark-diquark picture, the spin coupling between the diquark and the spectator quark. Although the magnetic moments of triply heavy baryons are not yet experimentally accessible, theoretical predictions are valuable for guiding future measurements and for discriminating among competing models of baryon structure.

The magnetic moments of triply heavy baryons are calculated within the constituent quark model by evaluating the contributions of individual quark magnetic moments weighted by the spin--flavor wave functions. The magnetic moment of a baryon with total angular momentum projection $M_J = J$ is given by
\begin{equation}
\mu_B = \sum_i \left\langle \phi_{sf} \left| \frac{e_i}{2 m_i^{\mathrm{eff}}} \, \sigma_{iz}\right| \phi_{sf} \right\rangle,
\label{eq:muB}
\end{equation}
where $\phi_{sf}$ denotes the spin--flavor wave function, $e_i$ is the electric charge of the $i$-th quark, $m_i^{\mathrm{eff}}$ is its effective mass, and $\sigma_{iz}$ is the $z$-component of the Pauli spin operator. Defining the magnetic moment of an individual quark as $\mu_i \equiv e_i/(2m_i^{\mathrm{eff}})$, Eq.~(\ref{eq:muB}) reduces to $\mu_B = \sum_i \mu_i \langle \sigma_{iz}\rangle$.

The effective quark mass accounts for binding effects within the baryon and is defined as
\begin{equation}
m_i^{\mathrm{eff}} = m_i \left(1 + \frac{\langle H \rangle}{\sum_j m_j} \right),
\label{eq:meff}
\end{equation}
where $\langle H \rangle$ is the expectation value of the interaction Hamiltonian in the baryon state.

To illustrate how the spin--flavor structure determines the magnetic moment, we work through the case of $\Omega_{ccb}^*$ ($J^P = 3/2^+$) explicitly. The two identical charm quarks in the ground state must form a spin-symmetric pair due to the Pauli principle (the color-antitriplet wave function is antisymmetric, the spatial ground state is symmetric, and the flavor wave function is symmetric under $c\leftrightarrow c$ exchange), so the $cc$ pair carries spin $S_{cc}=1$. Coupling $S_{cc}=1$ with the bottom quark spin $S_b=1/2$ to $J=3/2$, the fully stretched state $|J=3/2,\, M_J=3/2\rangle$ takes the form
\begin{equation}
    \left|\tfrac{3}{2},\,\tfrac{3}{2}\right\rangle = |1,1\rangle_{cc}\,\left|\tfrac{1}{2},\,\tfrac{1}{2}\right\rangle_b = |\!\uparrow\uparrow\rangle_{cc}\,|\!\uparrow\rangle_b,
\end{equation}
in which all three quark spins are aligned. The expectation values are therefore $\langle\sigma_{1z}\rangle = \langle\sigma_{2z}\rangle = \langle\sigma_{3z}\rangle = +1$, and the magnetic moment is
\begin{equation}
    \mu(\Omega_{ccb}^*) = \mu_c + \mu_c + \mu_b = 2\mu_c + \mu_b.
\end{equation}
For the remaining states, the same Clebsch--Gordan decomposition is applied with the appropriate total spin. The explicit magnetic moment expressions for the $\Omega_{ccb}$ and $\Omega_{bbc}$ systems are presented in the following section. Here, we note that, within SU(6) spin--flavor symmetry, the $J=3/2$ states receive aligned contributions from all three quarks, whereas the $J=1/2$ states reflect the anti-alignment of the unpaired quark spin relative to the diquark spin, leading to a partial cancellation between the charm and bottom quark moments.

\subsection{Regge trajectories}
\label{subsec:regge}

Regge theory, originally formulated in the context of complex angular momentum in scattering amplitudes~\cite{Regge:1959mz,Regge:1960zc}, has become a powerful phenomenological tool in hadron spectroscopy for organizing mass spectra and testing the internal consistency of quark models~\cite{Collins:1971ff,Collins:1977jy,Chew:1961ev}. A central prediction of the theory is that hadron states belonging to the same family lie on approximately linear trajectories when plotted in the $(J,\,M^2)$ or $(n_r,\,M^2)$ planes, where $J$ is the total angular momentum and $n_r$ the radial quantum number. This linear behavior has been well established for light mesons and baryons~\cite{Klempt:2012fy,Ebert:2009ub} and has been extended to the heavy-quark sector, including singly heavy~\cite{Ebert:2011kk,Chen:2018bbr,Jia:2024pyb}, doubly heavy~\cite{Oudichhya:2021yln}, and triply heavy baryons~\cite{Shah:2017liu,Shah:2018ont,Shah:2023zph,Oudichhya:2023lva,Xie:2024lfo}. The physical origin of the linearity can be understood from the string picture of confinement: the linear confining potential gives rise to a flux tube between the constituents whose rotational and vibrational excitations naturally produce trajectories that are linear in $M^2$~\cite{Nambu:1974zg,Selem:2006nd}.

In this work, we employ the Regge analysis in the $(n_r,\,M^2)$ plane, where $n_r$ denotes the radial excitation quantum number. This choice isolates the radial dynamics and avoids the complications associated with spin-orbit and tensor splittings that affect trajectories in the $(J,\,M^2)$ plane for states with $L\geq 1$. The radial Regge trajectory is parametrized as
\begin{equation}
    M^2_{n_r} = \beta_0 + \beta\, n_r,
    \label{eq:regge}
\end{equation}
where $\beta$ is the slope and $\beta_0$ the intercept. The slope $\beta$ is related to the string tension of the confining potential and is expected to be approximately universal for states within the same flavor sector~\cite{Ebert:2011kk,Chen:2023cws}, while the intercept $\beta_0$ encodes the ground-state mass and the short-distance dynamics.

For each quark-diquark channel of the $\Omega_{ccb}$ and $\Omega_{cbb}$ baryons, the calculated masses of the radially excited S-wave states are fitted to Eq.~(\ref{eq:regge}) using a least-squares procedure to extract $\beta$ and $\beta_0$. The quality of the linear fit is assessed through the coefficient of determination $R^2$~\footnote{The coefficient of determination $R^2$ is defined as $R^2 = 1 - \sum_i (y_i - y_i^{\mathrm{fit}})^2 / \sum_i (y_i - \bar{y})^2$, where $y_i$ are the calculated values, $y_i^{\mathrm{fit}}$ are the corresponding values from the fit, and $\bar{y}$ is the mean of the data. It measures how well the fitted function reproduces the data, with $R^2 = 1$ indicating a perfect fit.}. Obtaining $R^2$ values close to unity would confirm that the computed spectrum exhibits the regular, equidistant pattern in $M^2$ expected from the confining dynamics encoded in the Cornell potential.

The Regge analysis serves a twofold purpose. First, it provides a nontrivial self-consistency check on the model: a spectrum derived from the Cornell potential with linear confinement should, by construction, yield approximately linear radial trajectories, and any significant deviation would signal numerical artifacts or an inadequacy of the fitting procedure. Second, the fitted trajectories can be extrapolated to higher radial excitations ($n_r = 4,\,5,\ldots$) to predict masses of states that lie beyond the range explicitly computed from the Schr\"{o}dinger equation. Such extrapolated predictions, while subject to increasing uncertainty at high excitation numbers, provide useful benchmarks for comparison with other theoretical approaches~\cite{Shah:2017liu,Oudichhya:2023lva,Xie:2024lfo} and can guide the identification of triply heavy baryon candidates in future experiments.

\section{Numerical Results and Discussion}
\label{sec:numresdis}

\subsection{Mass Spectra}
\label{subsec:mass_spectra}

The mass spectra of the $\Omega_{ccb}$ and $\Omega_{bbc}$ baryons are obtained by numerically solving the radial Schr\"{o}dinger equation given in Eq.~(\ref{eq:SchrodingerRedQd2}) within the quark--diquark framework. The predicted mass spectra for the $\Omega_{ccb}$ and $\Omega_{bbc}$ baryons are presented in Tables~\ref{tab:Omega_ccb} and~\ref{tab:Omega_bbc}, respectively.

\begin{table}[h]
\centering
\caption{Predicted mass spectrum of $\Omega_{ccb}$ baryons for scalar and axial-vector diquark configurations (in GeV).}
\label{tab:Omega_ccb}
\begin{tabular}{cccccc}
\hline\hline
State & \multicolumn{2}{c}{$\Omega_{b\{cc\}}$} & $\Omega_{c[bc]}$ & \multicolumn{2}{c}{$\Omega_{c\{bc\}}$} \\
\cline{2-3} \cline{5-6}
 & $J^P=\tfrac{1}{2}^+$ & $J^P=\tfrac{3}{2}^+$ & $J^P=\tfrac{1}{2}^+$ & $J^P=\tfrac{1}{2}^+$ & $J^P=\tfrac{3}{2}^+$ \\
\hline
1S & 7.854 & 7.855 & 7.977 & 7.975 & 7.978 \\
2S & 8.462 & 8.463 & 8.575 & 8.574 & 8.576 \\
3S & 8.833 & 8.833 & 8.978 & 8.978 & 8.979 \\
4S & 9.134 & 9.134 & 9.314 & 9.314 & 9.315 \\
\hline
 & $J^P=\tfrac{1}{2}^-$ & $J^P=\tfrac{3}{2}^-$ & $J^P=\tfrac{1}{2}^-$ & $J^P=\tfrac{1}{2}^-$ & $J^P=\tfrac{3}{2}^-$ \\
\hline
1P & 8.353 & 8.354 & 8.436 & 8.435 & 8.437 \\
2P & 8.735 & 8.736 & 8.855 & 8.854 & 8.856 \\
3P & 9.044 & 9.044 & 9.201 & 9.200 & 9.201 \\
4P & 9.313 & 9.314 & 9.507 & 9.506 & 9.507 \\
\hline\hline
\end{tabular}
\end{table}

\begin{table}[h]
\centering
\caption{Predicted mass spectrum of $\Omega_{cbb}$ baryons for scalar and axial-vector diquark configurations (in GeV).}
\label{tab:Omega_bbc}
\begin{tabular}{cccccc}
\hline\hline
State & \multicolumn{2}{c}{$\Omega_{c\{bb\}}$} & $\Omega_{b[bc]}$ & \multicolumn{2}{c}{$\Omega_{b\{bc\}}$} \\
\cline{2-3} \cline{5-6}
 & $J^P=\tfrac{1}{2}^+$ & $J^P=\tfrac{3}{2}^+$ & $J^P=\tfrac{1}{2}^+$ & $J^P=\tfrac{1}{2}^+$ & $J^P=\tfrac{3}{2}^+$ \\
\hline
1S & 10.882 & 10.885 & 10.677 & 10.677 & 10.677 \\
2S & 11.480 & 11.482 & 11.320 & 11.320 & 11.321 \\
3S & 11.879 & 11.880 & 11.676 & 11.676 & 11.676 \\
4S & 12.210 & 12.210 & 11.957 & 11.957 & 11.957 \\
\hline
 & $J^P=\tfrac{1}{2}^-$ & $J^P=\tfrac{3}{2}^-$ & $J^P=\tfrac{1}{2}^-$ & $J^P=\tfrac{1}{2}^-$ & $J^P=\tfrac{3}{2}^-$ \\
\hline
1P & 11.346 & 11.348 & 11.232 & 11.231 & 11.232 \\
2P & 11.759 & 11.761 & 11.595 & 11.595 & 11.595 \\
3P & 12.099 & 12.100 & 11.882 & 11.881 & 11.882 \\
4P & 12.400 & 12.400 & 12.130 & 12.130 & 12.130 \\
\hline\hline
\end{tabular}
\end{table}

For the $\Omega_{ccb}$ ground state, the $\Omega_{b\{cc\}}$ configuration yields $7.854$~GeV ($J^P=\tfrac{1}{2}^+$), while $\Omega_{c[bc]}$ and $\Omega_{c\{bc\}}$ give $7.977$~GeV and $7.975$--$7.978$~GeV, respectively, producing a spread of approximately $124$~MeV. For the $\Omega_{cbb}$ ground state, $\Omega_{c\{bb\}}$ gives $10.882$~GeV, whereas both $\Omega_{b[bc]}$ and $\Omega_{b\{bc\}}$ yield $10.677$~GeV, resulting in a larger spread of $\sim 208$~MeV. The pattern is reversed between the two systems: the equal-flavor diquark configuration produces the lower mass for $\Omega_{ccb}$ but the higher mass for $\Omega_{cbb}$. This inversion is traced to the different reduced masses $\mu_{Qd}$ in each decomposition; in $\Omega_{c\{bb\}}$ the lighter charm quark paired with the heavy $\{bb\}$ diquark yields a smaller reduced mass and weaker binding, raising the total mass. These spreads remain approximately constant across all excitations, indicating that the systematic uncertainty of the two-body reduction does not grow with excitation energy. The larger spread for $\Omega_{bbc}$ reflects the greater mass asymmetry in that system.

A striking feature is the near-degeneracy of the scalar and axial-vector $bc$ diquark configurations. The $\Omega_{c[bc]}$ and $\Omega_{c\{bc\}}$ spectra differ by at most $3$~MeV for $\Omega_{ccb}$, while $\Omega_{b[bc]}$ and $\Omega_{b\{bc\}}$ are degenerate to below $1$~MeV for $\Omega_{bbc}$. This reflects the suppression of the chromomagnetic splitting within the $bc$ diquark due to the large reduced mass $\mu_{bc}\approx 1.1$~GeV, which renders the scalar and axial-vector diquark masses nearly identical.

The hyperfine splitting between the $J=1/2$ and $J=3/2$ states is strongly suppressed throughout both spectra. For the ground states, it ranges from $0$ to $3$~MeV and decreases further with increasing excitation, approaching zero for the highest computed states. The same suppression holds in the negative-parity sector, where the $J^P=\tfrac{1}{2}^-$ and $\tfrac{3}{2}^-$ P-wave states are split by at most $2$~MeV. This is a direct consequence of the $1/\mu_{Qd}^2$ dependence of the spin-spin interaction, which is heavily damped when both constituents are heavy.

The radial excitation energies exhibit a characteristic decreasing pattern. For $\Omega_{b\{cc\}}$, the successive S-wave spacings are $608$, $371$, and $301$~MeV; for $\Omega_{c\{bb\}}$, they are $598$, $399$, and $331$~MeV. This behavior is a well-known feature of the Cornell potential, where higher excitations increasingly probe the linear confining regime. The $2S$--$1S$ spacing of $\sim 600$~MeV is remarkably stable across both systems and all configurations. The $1P$ state lies between the $1S$ and $2S$ states in all cases, with a $1P$--$1S$ splitting ranging from $459$ to $555$~MeV, consistent with the level ordering expected from a potential with both Coulomb and linear terms.

These systematic features collectively demonstrate that the present quark--diquark framework provides a consistent and physically well-grounded description of triply heavy baryons. In particular, the stability of the radial excitation pattern, the suppression of hyperfine splittings, and the near-degeneracy of the $bc$ diquark configurations all emerge naturally from heavy-quark dynamics, without the need for additional model assumptions. This internal consistency indicates that the essential QCD-driven mechanisms---namely the interplay between short-range Coulombic attraction and long-range confinement---are effectively captured within the present approach. Moreover, the smooth behavior of the spectra across different diquark configurations and excitation levels suggests that the two-body reduction remains reliable even in highly excited states. These observations reinforce the predictive power of the model and support its applicability to unexplored regions of the triply heavy baryon spectrum, where future experimental or lattice QCD results may provide further validation.

\begin{table*}[htbp]
\centering
\caption{Comparison of mass spectra (in GeV) for $\Omega_{ccb}$ baryons.}
\label{tab:comparison_ccb}
\begin{ruledtabular}
\begin{tabular}{ccccccccccc}
Baryon & $J^P$ & $n$ & This work & \cite{Yang:2019lsg} & \cite{Faustov:2021qqf} & \cite{Yu:2025gdg} & \cite{Zhou:2025fpp} & \cite{Oudichhya:2023pkg} & \cite{Shah:2019jxp} & \cite{Qin:2019hgk} \\
\hline
\multirow{16}{*}{$\Omega_{\{cc\}b}$ ($s=1$)}
& $\tfrac{1}{2}^+$ & 1S & 7.854 & 8.004 & 7.984 & 8.025 & 8.017 & 8.192 & 8.005 & 7.867 \\
&                  & 2S & 8.462 & 8.455 & 8.361 & 8.422 & 8.463 & 8.621 & 8.606 & 8.337 \\
&                  & 3S & 8.833 & 8.536 & 8.405 & 8.522 & 8.605 & 9.030 & 9.067 &       \\
&                  & 4S & 9.134 &       &       & 8.731 & 8.825 & 9.420 & 9.491 &       \\
\cline{2-11}
& $\tfrac{1}{2}^-$ & 1P & 8.353 & 8.306 & 8.250 & 8.303 & 8.319 &       & 8.487 & 8.164 \\
&                  & 2P & 8.735 & 8.663 & 8.583 & 8.611 & 8.657 &       & 8.947 &       \\
&                  & 3P & 9.044 &       &       & 8.738 & 8.825 &       & 9.374 &       \\
&                  & 4P & 9.313 &       &       & 8.881 &       &       & 9.775 &       \\
\cline{2-11}
& $\tfrac{3}{2}^+$ & 1S & 7.855 & 8.023 & 7.999 & 8.046 & 8.030 & 8.223 & 8.049 & 7.963 \\
&                  & 2S & 8.463 & 8.468 & 8.366 & 8.438 & 8.469 & 8.637 & 8.624 & 8.427 \\
&                  & 3S & 8.833 & 8.536 & 8.412 & 8.563 & 8.603 & 9.032 & 9.076 &       \\
&                  & 4S & 9.134 &       &       & 8.745 &       & 9.410 & 9.495 &       \\
\cline{2-11}
& $\tfrac{3}{2}^-$ & 1P & 8.354 & 8.306 & 8.262 & 8.302 & 8.322 &       & 8.476 & 8.275 \\
&                  & 2P & 8.736 & 8.663 & 8.591 & 8.609 & 8.808 &       & 8.939 &       \\
&                  & 3P & 9.044 &       &       & 8.738 &       &       & 9.368 &       \\
&                  & 4P & 9.314 &       &       & 8.878 &       &       & 9.769 &       \\
\hline
\multirow{8}{*}{$\Omega_{c[cb]}$ ($s=0$)}
& $\tfrac{1}{2}^+$ & 1S & 7.977 & 8.004 & 7.984 & 8.025 & 8.017 & 8.192 & 8.005 & 7.867 \\
&                  & 2S & 8.575 & 8.455 & 8.361 & 8.422 & 8.463 & 8.621 & 8.606 & 8.337 \\
&                  & 3S & 8.978 & 8.536 & 8.405 & 8.522 & 8.605 & 9.030 & 9.067 &       \\
&                  & 4S & 9.314 &       &       & 8.731 & 8.825 & 9.420 & 9.491 &       \\
\cline{2-11}
& $\tfrac{1}{2}^-$ & 1P & 8.436 & 8.306 & 8.250 & 8.303 & 8.319 &       & 8.487 & 8.164 \\
&                  & 2P & 8.855 & 8.663 & 8.583 & 8.611 & 8.657 &       & 8.947 &       \\
&                  & 3P & 9.201 &       &       & 8.738 & 8.825 &       & 9.374 &       \\
&                  & 4P & 9.507 &       &       & 8.881 &       &       & 9.775 &       \\
\hline
\multirow{16}{*}{$\Omega_{c\{cb\}}$ ($s=1$)}
& $\tfrac{1}{2}^+$ & 1S & 7.975 & 8.004 & 7.984 & 8.025 & 8.017 & 8.192 & 8.005 & 7.867 \\
&                  & 2S & 8.574 & 8.455 & 8.361 & 8.422 & 8.463 & 8.621 & 8.606 & 8.337 \\
&                  & 3S & 8.978 & 8.536 & 8.405 & 8.522 & 8.605 & 9.030 & 9.067 &       \\
&                  & 4S & 9.314 &       &       & 8.731 & 8.825 & 9.420 & 9.491 &       \\
\cline{2-11}
& $\tfrac{1}{2}^-$ & 1P & 8.435 & 8.306 & 8.250 & 8.303 & 8.319 &       & 8.487 & 8.164 \\
&                  & 2P & 8.854 & 8.663 & 8.583 & 8.611 & 8.657 &       & 8.947 &       \\
&                  & 3P & 9.200 &       &       & 8.738 & 8.825 &       & 9.374 &       \\
&                  & 4P & 9.506 &       &       & 8.881 &       &       & 9.775 &       \\
\cline{2-11}
& $\tfrac{3}{2}^+$ & 1S & 7.978 & 8.023 & 7.999 & 8.046 & 8.030 & 8.223 & 8.049 & 7.963 \\
&                  & 2S & 8.576 & 8.468 & 8.366 & 8.438 & 8.469 & 8.637 & 8.624 & 8.427 \\
&                  & 3S & 8.979 & 8.536 & 8.412 & 8.563 & 8.603 & 9.032 & 9.076 &       \\
&                  & 4S & 9.315 &       &       & 8.745 &       & 9.410 & 9.495 &       \\
\cline{2-11}
& $\tfrac{3}{2}^-$ & 1P & 8.437 & 8.306 & 8.262 & 8.302 & 8.322 &       & 8.476 & 8.275 \\
&                  & 2P & 8.856 & 8.663 & 8.591 & 8.609 & 8.808 &       & 8.939 &       \\
&                  & 3P & 9.201 &       &       & 8.738 &       &       & 9.368 &       \\
&                  & 4P & 9.507 &       &       & 8.878 &       &       & 9.769 &       \\
\end{tabular}
\end{ruledtabular}
\end{table*}

The ground-state masses are compared with selected theoretical predictions in Table~\ref{tab:comparison_ccb} for the $\Omega_{ccb}$ baryon and in Table~\ref{tab:comparison_bbc} for the $\Omega_{cbb}$ baryon. Overall, the predicted masses are in reasonable agreement with previous studies, with relativistic approaches generally yielding slightly higher values, while the present nonrelativistic quark--diquark framework produces a somewhat more compact spectrum. A closer inspection, however, reveals a systematic pattern that depends on both the diquark configuration and the radial excitation level, which deserves detailed discussion.

\begin{table*}[htbp]
\centering
\caption{Comparison of mass spectra (in GeV) for $\Omega_{bbc}$ baryons.}
\label{tab:comparison_bbc}
\begin{ruledtabular}
\begin{tabular}{ccccccccccc}
Baryon & $J^P$ & $n$ & This work & \cite{Yang:2019lsg} & \cite{Faustov:2021qqf} & \cite{Yu:2025gdg} & \cite{Zhou:2025fpp} & \cite{Oudichhya:2023pkg} & \cite{Shah:2019jxp} & \cite{Qin:2019hgk} \\
\hline
\multirow{16}{*}{$\Omega_{c\{bb\}}$ ($s=1$)}
& $\tfrac{1}{2}^+$ & 1S & 10.882 & 11.200 & 11.198 & 11.217 & 11.204 & 11.526 & 11.198 & 11.077 \\
&                  & 2S & 11.480 & 11.607 & 11.507 & 11.604 & 11.621 & 11.757 & 11.749 & 11.603 \\
&                  & 3S & 11.879 & 11.677 & 11.622 & 11.700 &        & 11.984 & 12.146 &        \\
&                  & 4S & 12.210 &        &        & 11.888 &        & 12.206 & 12.511 &        \\
\cline{2-11}
& $\tfrac{1}{2}^-$ & 1P & 11.346 & 11.482 & 11.414 & 11.492 & 11.496 &        & 11.649 & 11.413 \\
&                  & 2P & 11.759 &        & 11.506 & 11.798 &        &        & 12.042 &        \\
&                  & 3P & 12.099 &        &        & 11.900 &        &        & 12.409 &        \\
&                  & 4P & 12.400 &        &        & 12.046 &        &        & 12.754 &        \\
\cline{2-11}
& $\tfrac{3}{2}^+$ & 1S & 10.885 & 11.221 & 11.217 & 11.236 &        & 11.541 & 11.217 & 11.167 \\
&                  & 2S & 11.482 & 11.622 & 11.515 & 11.617 &        & 11.779 & 11.773 & 11.703 \\
&                  & 3S & 11.880 & 11.677 & 11.629 & 11.709 &        & 12.012 & 12.157 &        \\
&                  & 4S & 12.210 &        &        & 11.899 &        & 12.241 & 12.517 &        \\
\cline{2-11}
& $\tfrac{3}{2}^-$ & 1P & 11.348 & 11.482 & 11.424 & 11.506 & 11.506 & 11.885 & 11.644 & 11.523 \\
&                  & 2P & 11.761 &        & 11.535 & 11.809 &        &        & 12.038 &        \\
&                  & 3P & 12.100 &        &        & 11.900 &        &        & 12.405 &        \\
&                  & 4P & 12.400 &        &        & 12.057 &        &        & 12.751 &        \\
\hline
\multirow{8}{*}{$\Omega_{b[bc]}$ ($s=0$)}
& $\tfrac{1}{2}^+$ & 1S & 10.677 & 11.200 & 11.198 & 11.217 & 11.204 & 11.526 & 11.198 & 11.077 \\
&                  & 2S & 11.320 & 11.607 & 11.507 & 11.604 & 11.621 & 11.757 & 11.749 & 11.603 \\
&                  & 3S & 11.676 & 11.677 & 11.622 & 11.700 &        & 11.984 & 12.146 &        \\
&                  & 4S & 11.957 &        &        & 11.888 &        & 12.206 & 12.511 &        \\
\cline{2-11}
& $\tfrac{1}{2}^-$ & 1P & 11.232 & 11.482 & 11.414 & 11.492 & 11.496 &        & 11.649 & 11.413 \\
&                  & 2P & 11.595 &        & 11.506 & 11.798 &        &        & 12.042 &        \\
&                  & 3P & 11.882 &        &        & 11.900 &        &        & 12.409 &        \\
&                  & 4P & 12.130 &        &        & 12.046 &        &        & 12.754 &        \\
\hline
\multirow{16}{*}{$\Omega_{b\{bc\}}$ ($s=1$)}
& $\tfrac{1}{2}^+$ & 1S & 10.677 & 11.200 & 11.198 & 11.217 & 11.204 & 11.526 & 11.198 & 11.077 \\
&                  & 2S & 11.320 & 11.607 & 11.507 & 11.604 & 11.621 & 11.757 & 11.749 & 11.603 \\
&                  & 3S & 11.676 & 11.677 & 11.622 & 11.700 &        & 11.984 & 12.146 &        \\
&                  & 4S & 11.957 &        &        & 11.888 &        & 12.206 & 12.511 &        \\
\cline{2-11}
& $\tfrac{1}{2}^-$ & 1P & 11.231 & 11.482 & 11.414 & 11.492 & 11.496 &        & 11.649 & 11.413 \\
&                  & 2P & 11.595 &        & 11.506 & 11.798 &        &        & 12.042 &        \\
&                  & 3P & 11.881 &        &        & 11.900 &        &        & 12.409 &        \\
&                  & 4P & 12.130 &        &        & 12.046 &        &        & 12.754 &        \\
\cline{2-11}
& $\tfrac{3}{2}^+$ & 1S & 10.677 & 11.221 & 11.217 & 11.236 &        & 11.541 & 11.217 & 11.167 \\
&                  & 2S & 11.321 & 11.622 & 11.515 & 11.617 &        & 11.779 & 11.773 & 11.703 \\
&                  & 3S & 11.676 & 11.677 & 11.629 & 11.709 &        & 12.012 & 12.157 &        \\
&                  & 4S & 11.957 &        &        & 11.899 &        & 12.241 & 12.517 &        \\
\cline{2-11}
& $\tfrac{3}{2}^-$ & 1P & 11.232 & 11.482 & 11.424 & 11.506 & 11.506 & 11.885 & 11.644 & 11.523 \\
&                  & 2P & 11.595 &        & 11.535 & 11.809 &        &        & 12.038 &        \\
&                  & 3P & 11.882 &        &        & 11.900 &        &        & 12.405 &        \\
&                  & 4P & 12.130 &        &        & 12.057 &        &        & 12.751 &        \\
\end{tabular}
\end{ruledtabular}
\end{table*}

For the $\Omega_{ccb}$ sector (Table~\ref{tab:comparison_ccb}), the ground-state $1S$ mass of the $\Omega_{b\{cc\}}$ configuration, $7.854$~GeV, lies roughly $150$~MeV below the median of the compared predictions ($\sim\!8.01$~GeV), while the mixed-flavor clusterings $\Omega_{c[cb]}$ and $\Omega_{c\{cb\}}$ both yield $\simeq 7.98$~GeV, in close agreement with the literature consensus. A similar pattern persists in the radial excitations: the $\Omega_{c[cb]}$ and $\Omega_{c\{cb\}}$ masses remain within $50$--$150$~MeV of the cited predictions throughout the $2S$--$4S$ sequence, whereas the $\Omega_{b\{cc\}}$ spectrum is consistently the lowest of the three clusterings. On this basis, the mixed-flavor $bc$ diquark configurations emerge as the more favored quark--diquark decomposition of $\Omega_{ccb}$ when compared to the broader theoretical consensus.

In the $\Omega_{cbb}$ sector (Table~\ref{tab:comparison_bbc}), the ordering of configurations is reversed. The $\Omega_{c\{bb\}}$ ground state at $10.882$~GeV lies approximately $320$~MeV below the literature median ($\sim\!11.20$~GeV), yet it is consistently the closest of our three clusterings to the cited predictions. The $\Omega_{b[bc]}$ and $\Omega_{b\{bc\}}$ configurations, degenerate at $10.677$~GeV, are shifted downward by a further $\sim 200$~MeV and thus deviate by $\sim\!520$~MeV from the literature median. Therefore, for $\Omega_{bbc}$ the equal-flavor $\{bb\}$ diquark paired with a charm spectator provides the favored decomposition.

Taken together, the two sectors point to a common structural feature: in both cases, the configuration closest to the literature consensus is the one in which a charm quark acts as the spectator of a more compact heavy diquark ($\{bc\}/[bc]$ in $\Omega_{ccb}$, $\{bb\}$ in $\Omega_{bbc}$). This pattern is consistent with expectations from heavy-quark scale separation: the tighter the internal diquark correlation, the more reliable the two-body reduction becomes, and the resulting quark--diquark system most closely mirrors the three-body dynamics captured by alternative approaches such as hypercentral, relativistic, or Dyson--Schwinger-based frameworks.

An additional observation is that the agreement with the literature improves systematically at higher excitations. For the $\Omega_{c\{bb\}}$ channel, the offset from the literature median decreases from $\sim\!320$~MeV at $1S$ to $\lesssim 100$~MeV by $3S$, and the $4S$ masses agree with the predictions of Refs.~\cite{Yu:2025gdg} and \cite{Oudichhya:2023pkg} within $10$--$20$~MeV. A similar convergence is observed for $\Omega_{c[cb]}/\Omega_{c\{cb\}}$ in the $\Omega_{ccb}$ sector. This behavior indicates that the systematic downward shift in the ground states originates primarily from the short-distance Coulombic dynamics, where differences in the effective strong coupling and in the treatment of relativistic corrections between models are most pronounced, whereas the long-range confinement that dominates the excited-state spacings is captured similarly across approaches.

A further noteworthy feature, already anticipated in the structure of Table~\ref{tab:comparison_bbc}, is the near-exact degeneracy of the scalar $[bc]$ and axial-vector $\{bc\}$ diquark configurations: the corresponding $\Omega_{b[bc]}$ and $\Omega_{b\{bc\}}$ spectra agree to within a few MeV at every level. This reflects the strong suppression of the chromomagnetic splitting inside the mixed-flavor $bc$ diquark, consistent with its large reduced mass $\mu_{bc} \approx 1.1$~GeV, and implies that the spin of the $bc$ diquark plays only a marginal role in determining the $\Omega_{bbc}$ spectrum.

The deviations between the present results and the cited predictions remain approximately uniform across radial and orbital excitations, indicating that the differences are systematic rather than state-dependent. This supports the internal robustness of the quark--diquark approximation, identifies the $\Omega_{b[bc]}/\Omega_{b\{bc\}}$ channels as equivalent representations within our framework, and confirms that the equal-flavor clustering ($\{bb\} + c$ for $\Omega_{bbc}$ and, symmetrically, the mixed-flavor clustering $\{bc\}/[bc] + c$ for $\Omega_{ccb}$) provides the most reliable quark--diquark decomposition when benchmarked against the existing theoretical literature.

\subsection{Magnetic moments}
\label{subsec:magmoments}

The magnetic moments of the triply heavy baryons $\Omega_{ccb}$, $\Omega_{ccb}^{*}$, $\Omega_{bbc}$, and $\Omega_{bbc}^{*}$ are evaluated within the constituent quark model using the standard nonrelativistic expression for the baryon magnetic moment,
\begin{equation}
\mu_B = \sum_{i}\langle\phi_{sf}|\,\hat{\mu}_{z,i}\,|\phi_{sf}\rangle,
\qquad
\hat{\mu}_{z,i} = \frac{e_i}{2m_i}\,\sigma_{z,i},
\end{equation}
where the sum runs over the three constituent quarks, $e_i$ and $m_i$ denote the electric charge and effective mass of the $i$-th quark, and $|\phi_{sf}\rangle$ is the spin--flavor wave function of the baryon state in question. The resulting analytical expressions, derived by sandwiching $\hat{\mu}_{z,i}$ between the spin--flavor eigenstates of the $J^{P}=\tfrac{1}{2}^{+}$ and $J^{P}=\tfrac{3}{2}^{+}$ configurations, are collected in Table~\ref{tab:magexp}.

\begin{table}[h]
\centering
\caption{Analytical expressions for the magnetic moments of the triply heavy baryons $\Omega_{ccb}^{(*)}$ and $\Omega_{bbc}^{(*)}$ in terms of the individual quark magnetic moments $\mu_Q = e_Q/(2m_Q)$.}
\label{tab:magexp}
\renewcommand{\arraystretch}{1.15}
\begin{tabular}{ccc}
\hline\hline
Baryon & $J^{P}$ & Magnetic moment \\
\hline
$\Omega_{ccb}$      & $\tfrac{1}{2}^{+}$ & $\tfrac{4}{3}\mu_c - \tfrac{1}{3}\mu_b$ \\[4pt]
$\Omega_{ccb}^{*}$  & $\tfrac{3}{2}^{+}$ & $2\mu_c + \mu_b$ \\[4pt]
$\Omega_{bbc}$      & $\tfrac{1}{2}^{+}$ & $\tfrac{4}{3}\mu_b - \tfrac{1}{3}\mu_c$ \\[4pt]
$\Omega_{bbc}^{*}$  & $\tfrac{3}{2}^{+}$ & $2\mu_b + \mu_c$ \\
\hline\hline
\end{tabular}
\end{table}

The structure of these expressions encodes the spin alignment of the constituent quarks in a transparent way. In the $J^{P}=\tfrac{1}{2}^{+}$ states, the two identical heavy quarks form a spin-$1$ pair whose third component is partially cancelled by the anti-aligned third quark, producing the characteristic $4/3$ and $-1/3$ Clebsch--Gordan coefficients. In the $J^{P}=\tfrac{3}{2}^{+}$ states, all three quark spins are fully aligned and contribute constructively; the magnetic moment reduces to the simple sum $2\mu_{Q_1}+\mu_{Q_2}$, with each quark weighted by its multiplicity. Two immediate consequences follow from the analytical forms alone, independently of the numerical values of $m_c$ and $m_b$. First, the $\Omega_{bbc}$ ($\tfrac{1}{2}^{+}$) magnetic moment is necessarily negative, because the dominant contribution $\tfrac{4}{3}\mu_b$ carries the negative sign of the $b$-quark charge, which is only partially offset by the positive $-\tfrac{1}{3}\mu_c$ subtraction; upon going to the fully aligned $\Omega_{bbc}^{*}$ ($\tfrac{3}{2}^{+}$), the positive $\mu_c$ now enters with full weight and drives the total moment positive, producing a distinctive sign flip that would unambiguously distinguish the ground state from its $\tfrac{3}{2}^{+}$ excitation in a future measurement. Second, a systematic enhancement is expected from $\tfrac{1}{2}^{+}$ to $\tfrac{3}{2}^{+}$ in both flavor sectors, because the partial cancellation present in the mixed-symmetry spin--flavor wave function of the lower-spin state is absent in the fully symmetric configuration.

\begin{table}[h]
\centering
\caption{Predicted magnetic moments (in nuclear magnetons $\mu_N$) of the triply heavy baryons for the ground-state $J^{P}=\tfrac{1}{2}^{+}$ and $J^{P}=\tfrac{3}{2}^{+}$ configurations.}
\label{tab:mm}
\renewcommand{\arraystretch}{1.15}
\begin{tabular}{ccc}
\hline\hline
Baryon & $J^{P}$ & $\mu\ [\mu_N]$ \\
\hline
$\Omega_{ccb}$      & $\tfrac{1}{2}^{+}$ & $\phantom{-}0.544$ \\
$\Omega_{ccb}^{*}$  & $\tfrac{3}{2}^{+}$ & $\phantom{-}0.718$ \\
$\Omega_{bbc}$      & $\tfrac{1}{2}^{+}$ & $-0.227$ \\
$\Omega_{bbc}^{*}$  & $\tfrac{3}{2}^{+}$ & $\phantom{-}0.267$ \\
\hline\hline
\end{tabular}
\end{table}

The numerical results are presented in Table~\ref{tab:mm}. The overall magnitudes are small---all below $1.0\,\mu_N$---as a direct consequence of the $1/m_Q$ suppression intrinsic to the heavy-quark magnetic moment: heavy quarks are intrinsically weak magnetic dipoles. The expected enhancement from $\tfrac{1}{2}^{+}$ to $\tfrac{3}{2}^{+}$ is clearly realized in both sectors, although its character differs markedly between the two flavor combinations. For the $\Omega_{ccb}^{(*)}$ doublet, the moment rises from $0.544\,\mu_N$ in the spin-$\tfrac{1}{2}$ ground state to $0.718\,\mu_N$ in its spin-aligned partner, an absolute shift of about $0.17\,\mu_N$ corresponding to a ratio of roughly $1.3$. For the $\Omega_{bbc}^{(*)}$ doublet, the transition is more pronounced: the moment changes from $-0.227\,\mu_N$ to $+0.267\,\mu_N$, reversing sign and moving by nearly $0.5\,\mu_N$ overall.

The sign pattern---positive for $\Omega_{ccb}$ and $\Omega_{ccb}^{*}$, negative for $\Omega_{bbc}$ but positive for $\Omega_{bbc}^{*}$---is in exact agreement with the analytical prediction discussed above and reflects the interplay between the individual quark magnetic moments and the spin--flavor Clebsch--Gordan coefficients that weight them. The results in Table~\ref{tab:mm} obey the magnitude hierarchy $|\mu(\Omega_{ccb}^{*})| > |\mu(\Omega_{ccb})| > |\mu(\Omega_{bbc}^{*})| > |\mu(\Omega_{bbc})|$, which is a model-independent consequence of the relative magnitudes of the quark magnetic moments $|\mu_c| > |\mu_b|$, ultimately controlled by the mass hierarchy $m_b > m_c$. This ordering should hold in any framework that respects the nonrelativistic limit.

To place these results in the context of the existing literature, we compare our predictions with a representative selection of theoretical estimates in Table~\ref{tab:mm_comparison}.

\begin{table*}[htbp]
\centering
\caption{Comparison of the predicted magnetic moments (in $\mu_N$) with selected theoretical results from the literature.}
\label{tab:mm_comparison}
\renewcommand{\arraystretch}{1.15}
\begin{ruledtabular}
\begin{tabular}{ccccccccc}
Baryon & This work & \cite{Faustov:2021qqf} & \cite{Oudichhya:2023pkg} & \cite{Shah:2023zph} & \cite{Mutuk:2021zes} & \cite{Hazra:2021lpa} & \cite{Thakkar:2016sog} & \cite{Faessler:2006ft} \\
\hline
$\Omega_{ccb}$      & $\phantom{-}0.544$  & $\phantom{-}0.609$  & $\phantom{-}0.563$  & $\phantom{-}0.565$  & $\phantom{-}0.540$  & $\phantom{-}0.526$  & $\phantom{-}0.526$  & $\phantom{-}0.502$ \\
$\Omega_{ccb}^{*}$  & $\phantom{-}0.718$  & $\phantom{-}0.819$  & $\phantom{-}0.750$  & $\phantom{-}0.751$  & $\phantom{-}0.720$  & $\phantom{-}0.695$  & $\phantom{-}0.695$  & $\phantom{-}0.651$ \\
$\Omega_{bbc}$      & $-0.227$ & $-0.239$ & $-0.222$ & $-0.223$ & $-0.210$ & $-0.210$ & $-0.209$ & $-0.203$ \\
$\Omega_{bbc}^{*}$  & $\phantom{-}0.267$  & $\phantom{-}0.335$  & $\phantom{-}0.290$  & $\phantom{-}0.285$  & $\phantom{-}0.270$  & $\phantom{-}0.256$  & $\phantom{-}0.255$  & $\phantom{-}0.216$ \\
\end{tabular}
\end{ruledtabular}
\end{table*}

The level of agreement across the literature is gratifying. For the $\Omega_{ccb}$ ground state, the seven cited calculations span $0.502$--$0.609\,\mu_N$, within which our value of $0.544\,\mu_N$ lies close to the median. An analogous situation holds for $\Omega_{ccb}^{*}$, where the literature range is $0.651$--$0.819\,\mu_N$ and our prediction of $0.718\,\mu_N$ again sits near the middle of the distribution. The negative-moment sector behaves similarly: our $\Omega_{bbc}$ value of $-0.227\,\mu_N$ is contained within the interval $-0.239$ to $-0.203\,\mu_N$ spanned by the other calculations, and our $\Omega_{bbc}^{*}$ prediction of $0.267\,\mu_N$ lies comfortably between the extremes $0.216\,\mu_N$ and $0.335\,\mu_N$. In all four channels, the dispersion between our values and the closest literature estimate is only a few hundredths of a nuclear magneton. Both the sign structure and the magnitude hierarchy identified above are consistently reproduced by every approach, confirming that these features are robust consequences of the heavy-quark dynamics rather than artifacts of any particular model.

A systematic trend is nonetheless visible in the comparison. The relativistic quark-model results of Ref.~\cite{Faustov:2021qqf} yield the largest values in every channel, while the nonrelativistic quark--diquark frameworks employed in Refs.~\cite{Hazra:2021lpa,Thakkar:2016sog,Faessler:2006ft} cluster at the lower end. This ordering is physically expected: relativistic corrections enhance the lower components of the quark spinors and effectively reduce the effective heavy-quark mass entering the magnetic moment, producing larger values, while diquark-based models partially absorb the quark substructure into the diquark mass, yielding a more compact bound-state description and correspondingly smaller electromagnetic couplings. The present results, together with those of Refs.~\cite{Oudichhya:2023pkg,Shah:2023zph,Mutuk:2021zes}, fall in the intermediate region between these two limiting behaviors, consistent with a nonrelativistic constituent-quark treatment that does not invoke diquark clustering but also does not include explicit Lorentz-boost corrections.

The residual spread across the literature---of the order of a few tenths of a nuclear magneton in each channel---can be traced to three main sources of model dependence: the choice of effective quark masses, which enter the magnetic moment inversely and can differ by several hundred MeV between parameter sets; the radial wave functions, whose value at the origin or overlap integrals modulate the matrix element in approaches that go beyond the static limit; and the inclusion or omission of relativistic, exchange-current, and anomalous-moment corrections.

Despite these differences, the overall consistency of the predictions across a broad range of methodologies---from relativistic quark models to nonrelativistic potential models, hypercentral approaches, and quark--diquark frameworks---provides strong evidence that the magnetic properties of triply heavy baryons are well constrained by the underlying heavy-quark dynamics and constitute robust, parameter-insensitive observables awaiting experimental confirmation.

\subsection{Regge Trajectories}
\label{subsec:reggeres}

In this subsection, we analyze the radial Regge behavior of the six quark--diquark decompositions studied in this work: $\Omega_{b\{cc\}}$, $\Omega_{c[bc]}$, and $\Omega_{c\{bc\}}$ in the $\Omega_{ccb}$ sector, and $\Omega_{c\{bb\}}$, $\Omega_{b[bc]}$, and $\Omega_{b\{bc\}}$ in the $\Omega_{bbc}$ sector. The objective is twofold: to test whether the computed spectrum conforms to the expected relation between the radial quantum number $n_r$ and the squared mass $M^{2}$, and to extract slope and intercept parameters that can serve as benchmark values for comparison with other theoretical approaches.

\begin{figure*}[htbp]
\centering
\begin{minipage}[t]{0.48\linewidth}
    \centering
    \includegraphics[width=\linewidth]{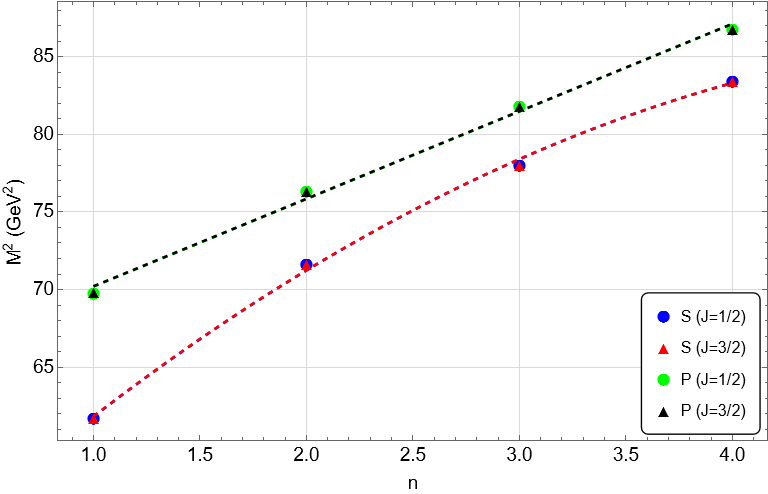}
    \caption{Radial Regge trajectory of the $\Omega_{b\{cc\}}$ baryon in the $(n_r, M^{2})$ plane.}
    \label{fig:regge_ccb1}
\end{minipage}
\hfill
\begin{minipage}[t]{0.48\linewidth}
    \centering
    \includegraphics[width=\linewidth]{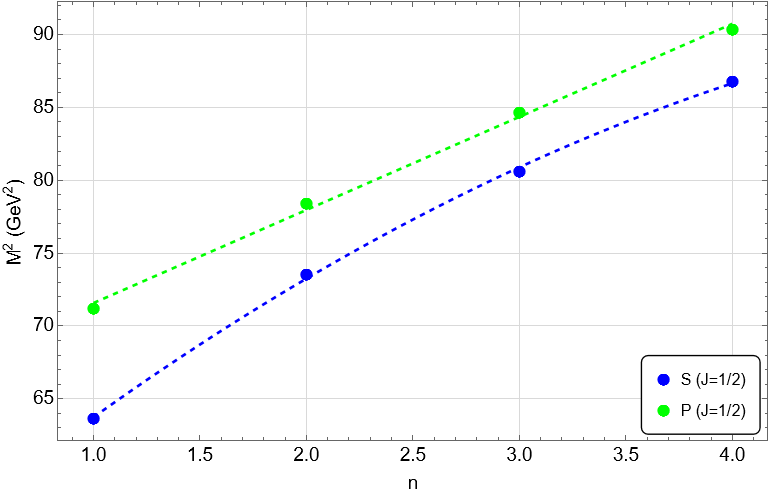}
    \caption{Radial Regge trajectory of the $\Omega_{c[bc]}$ baryon in the $(n_r, M^{2})$ plane.}
    \label{fig:regge_ccb2}
\end{minipage}
\end{figure*}

\begin{figure*}[htbp]
\centering
\begin{minipage}[t]{0.48\linewidth}
    \centering
    \includegraphics[width=\linewidth]{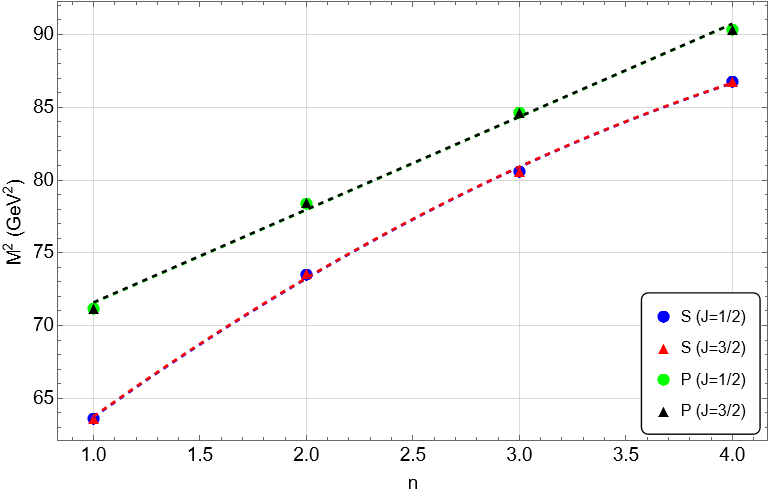}
    \caption{Radial Regge trajectory of the $\Omega_{c\{bc\}}$ baryon in the $(n_r, M^{2})$ plane.}
    \label{fig:regge_ccb3}
\end{minipage}
\hfill
\begin{minipage}[t]{0.48\linewidth}
    \centering
    \includegraphics[width=\linewidth]{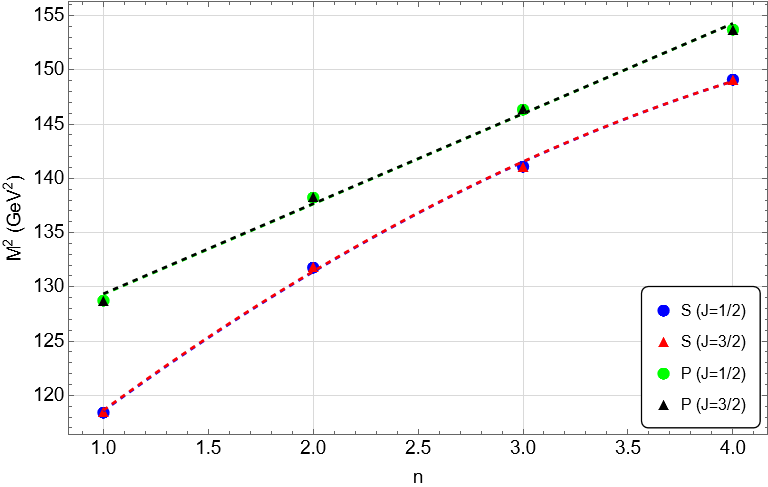}
    \caption{Radial Regge trajectory of the $\Omega_{c\{bb\}}$ baryon in the $(n_r, M^{2})$ plane.}
    \label{fig:regge_bbc1}
\end{minipage}
\end{figure*}

\begin{figure*}[htbp]
\centering
\begin{minipage}[t]{0.48\linewidth}
    \centering
    \includegraphics[width=\linewidth]{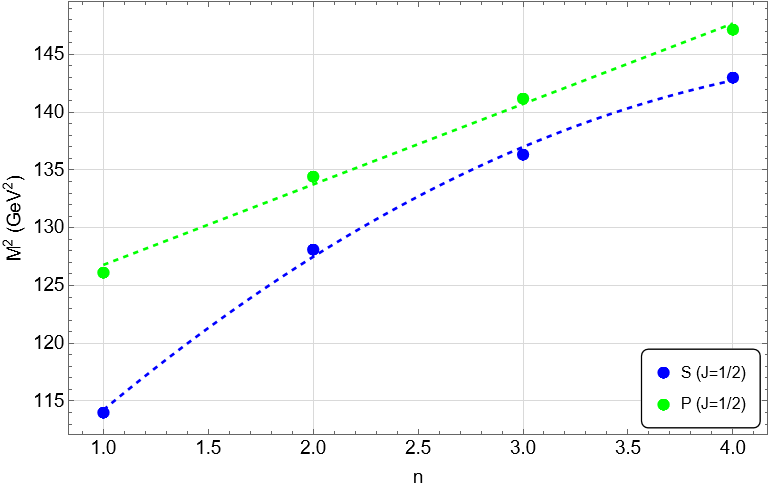}
    \caption{Radial Regge trajectory of the $\Omega_{b[bc]}$ baryon in the $(n_r, M^{2})$ plane.}
    \label{fig:regge_bbc3}
\end{minipage}
\hfill
\begin{minipage}[t]{0.48\linewidth}
    \centering
    \includegraphics[width=\linewidth]{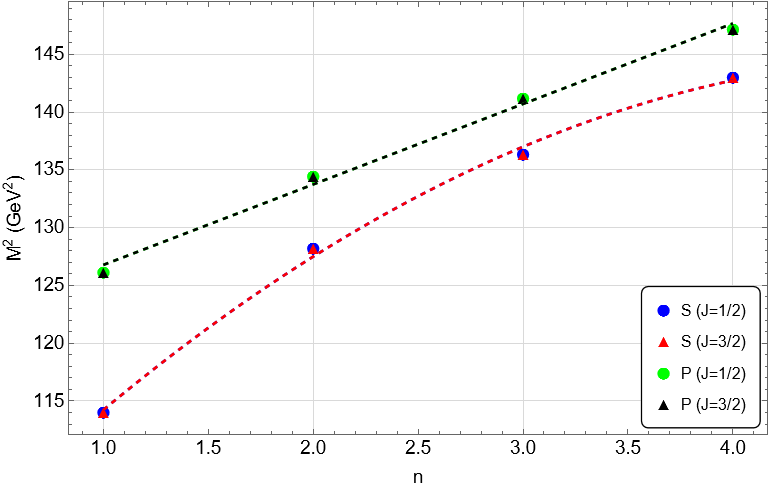}
    \caption{Radial Regge trajectory of the $\Omega_{b\{bc\}}$ baryon in the $(n_r, M^{2})$ plane.}
    \label{fig:regge_bbc2}
\end{minipage}
\end{figure*}

The trajectories are constructed in the $(n_r, M^{2})$ plane, with the $1S$--$4S$ and $1P$--$4P$ sequences fitted separately for each channel and for each total angular momentum assignment. The assumed functional form is given in Eq.~(\ref{eq:regge}), from which the slope $\beta$ and intercept $\beta_0$ are extracted by least squares, with the quality of the fit quantified by the coefficient of determination $R^{2}$. The resulting trajectories are displayed in Figs.~\ref{fig:regge_ccb1}--\ref{fig:regge_bbc2}, and the extracted parameters are collected in Tables~\ref{tab:regge_ccb} and~\ref{tab:regge_bbc}.

\begin{table}[htbp]
\centering
\caption{Slope ($\beta$), intercept ($\beta_0$), and coefficient of determination ($R^{2}$) of the radial Regge trajectories for the $\Omega_{ccb}$ baryon. Slopes and intercepts are given in units of $\mathrm{GeV}^{2}$.}
\label{tab:regge_ccb}
\begin{ruledtabular}
\begin{tabular}{ccccccc}
Baryon & Diquark & $J^P$ & $n$ & $\beta$ & $\beta_0$ & $R^{2}$ \\
\hline
\multirow{4}{*}{$\Omega_{\{cc\}b}$} & \multirow{4}{*}{$s=1$}
& $\tfrac{1}{2}^+$ & 1S--4S & 7.164 & 55.775 & 0.980 \\
& & $\tfrac{3}{2}^+$ & 1S--4S & 7.158 & 55.798 & 0.980 \\
& & $\tfrac{1}{2}^-$ & 1P--4P & 5.637 & 64.556 & 0.996 \\
& & $\tfrac{3}{2}^-$ & 1P--4P & 5.635 & 64.574 & 0.997 \\
\hline
\multirow{2}{*}{$\Omega_{c[bc]}$} & \multirow{2}{*}{$s=0$}
& $\tfrac{1}{2}^+$ & 1S--4S & 7.642 & 57.024 & 0.988 \\
& & $\tfrac{1}{2}^-$ & 1P--4P & 6.389 & 65.181 & 0.994 \\
\hline
\multirow{4}{*}{$\Omega_{c\{bc\}}$} & \multirow{4}{*}{$s=1$}
& $\tfrac{1}{2}^+$ & 1S--4S & 7.654 & 56.981 & 0.988 \\
& & $\tfrac{3}{2}^+$ & 1S--4S & 7.644 & 57.036 & 0.988 \\
& & $\tfrac{1}{2}^-$ & 1P--4P & 6.389 & 65.164 & 0.998 \\
& & $\tfrac{3}{2}^-$ & 1P--4P & 6.383 & 65.205 & 0.998 \\
\end{tabular}
\end{ruledtabular}
\end{table}

\begin{table}[htbp]
\centering
\caption{Slope ($\beta$), intercept ($\beta_0$), and coefficient of determination ($R^{2}$) of the radial Regge trajectories for the $\Omega_{bbc}$ baryon. Slopes and intercepts are given in units of $\mathrm{GeV}^{2}$.}
\label{tab:regge_bbc}
\begin{ruledtabular}
\begin{tabular}{ccccccc}
Baryon & Diquark & $J^P$ & $n$ & $\beta$ & $\beta_0$ & $R^{2}$ \\
\hline
\multirow{4}{*}{$\Omega_{c\{bb\}}$} & \multirow{4}{*}{$s=1$}
& $\tfrac{1}{2}^+$ & 1S--4S & 10.132 & 109.770 & 0.986 \\
& & $\tfrac{3}{2}^+$ & 1S--4S & 10.110 & 109.859 & 0.986 \\
& & $\tfrac{1}{2}^-$ & 1P--4P & 8.319 & 120.990 & 0.999 \\
& & $\tfrac{3}{2}^-$ & 1P--4P & 8.303 & 121.060 & 0.999 \\
\hline
\multirow{2}{*}{$\Omega_{b[bc]}$} & \multirow{2}{*}{$s=0$}
& $\tfrac{1}{2}^+$ & 1S--4S & 9.509 & 106.587 & 0.968 \\
& & $\tfrac{1}{2}^-$ & 1P--4P & 6.967 & 119.812 & 0.995 \\
\hline
\multirow{4}{*}{$\Omega_{b\{bc\}}$} & \multirow{4}{*}{$s=1$}
& $\tfrac{1}{2}^+$ & 1S--4S & 9.507 & 106.598 & 0.968 \\
& & $\tfrac{3}{2}^+$ & 1S--4S & 9.507 & 106.598 & 0.968 \\
& & $\tfrac{1}{2}^-$ & 1P--4P & 6.965 & 119.811 & 0.995 \\
& & $\tfrac{3}{2}^-$ & 1P--4P & 6.965 & 119.811 & 0.995 \\
\end{tabular}
\end{ruledtabular}
\end{table}

A first observation concerns the quality of the linear fits. The $P$-wave trajectories are accurately described by Eq.~(\ref{eq:regge}) in every channel, with $R^{2}$ values ranging from $0.994$ to $0.999$. The $S$-wave trajectories are of systematically lower quality, with $R^{2}$ between $0.968$ and $0.988$, the largest deviations from linearity occurring at the $1S$ level. This asymmetry is not a numerical artifact but a physical consequence of the Cornell potential: the $1S$ state probes a regime in which the Coulombic and linear contributions are comparable, whereas the excited states increasingly sample the asymptotically linear confining regime. The residual curvature observed for the $S$-waves is consistent with the $M^{2} \sim n_{r}^{4/3}$ scaling expected from confining dynamics and reported in alternative parametrizations of heavy-baryon Regge trajectories~\cite{Feng:2023txx}. The fact that this curvature is less pronounced for the $P$-waves, where the centrifugal barrier already suppresses the short-range Coulombic contribution, reinforces the interpretation that the nonlinearity is dynamical in origin rather than a purely kinematical feature of the parametrization.

Turning to the slope parameter, the extracted values span $\beta \simeq 5.6$--$10.1~\mathrm{GeV}^{2}$ across all channels, in line with the range reported in previous studies of heavy baryon Regge trajectories. Three systematic features emerge from Tables~\ref{tab:regge_ccb} and~\ref{tab:regge_bbc}. First, $\beta$ increases with the overall mass of the baryon: the $\Omega_{ccb}$ channels yield $\beta \simeq 5.6$--$7.7~\mathrm{GeV}^{2}$, while the $\Omega_{bbc}$ channels yield systematically larger values, $\beta \simeq 7.0$--$10.1~\mathrm{GeV}^{2}$. This positive correlation between slope and constituent mass is a direct consequence of heavy-quark dynamics within the Cornell potential. Second, within each family the $S$-wave slopes exceed the $P$-wave slopes by roughly $1.5$--$2~\mathrm{GeV}^{2}$. This hierarchy can be traced to the same mechanism that lowers the $S$-wave fit quality: the Coulombic depression of the $1S$ mass pulls the left endpoint of the trajectory downward, steepening the effective slope of a linear fit. The $P$-wave trajectories, free of this Coulombic pull on their ground state, sample the confining regime more uniformly and provide the more reliable slope estimates. Third, the slope is essentially independent of the spin of the mixed-flavor diquark: the $\Omega_{c[bc]}$ and $\Omega_{c\{bc\}}$ values agree to within $0.01~\mathrm{GeV}^{2}$, and the $\Omega_{b[bc]}$ and $\Omega_{b\{bc\}}$ values are degenerate to three significant figures. This is a direct echo of the strong suppression of the chromomagnetic splitting in the $bc$ diquark noted in Sec.~\ref{subsec:mass_spectra}, and implies that the Regge slope is governed by the overall binding rather than by the fine-structure content of the diquark.

The intercept $\beta_0$, corresponding to the extrapolated squared mass at $n_r = 0$, separates cleanly into two regimes set by the heavy-quark content. For the $\Omega_{ccb}$ channels, $\beta_0 \simeq 55.8$--$65.2~\mathrm{GeV}^{2}$, while for the $\Omega_{bbc}$ channels $\beta_0 \simeq 106.6$--$121.1~\mathrm{GeV}^{2}$. Taking the equal-flavor configurations as reference points, $\sqrt{\beta_0}$ yields $\simeq 7.47$~GeV for $\Omega_{b\{cc\}}$ and $\simeq 10.48$~GeV for $\Omega_{c\{bb\}}$, values that reproduce the ratio of the physical ground-state masses ($M_{\Omega_{bbc}}/M_{\Omega_{ccb}} \approx 1.39$) to better than $1\%$ accuracy. The intercept is therefore driven primarily by the heavy-quark rest-mass content of each baryon rather than by dynamical binding effects. A secondary trend, consistent across all channels, is that $\beta_0$ is slightly larger for the $P$-wave trajectories than for the $S$-wave ones, reflecting the larger effective size of orbitally excited states.

The dependence on the diquark composition, though subleading to the overall mass scale, is nevertheless visible and, interestingly, reverses between the two sectors. In the $\Omega_{ccb}$ sector, the equal-flavor configuration $\Omega_{b\{cc\}}$ yields both the smallest slope ($\beta \simeq 7.16$) and the smallest intercept ($\beta_0 \simeq 55.8$) among the three clusterings, whereas the mixed-flavor configurations $\Omega_{c[bc]}$ and $\Omega_{c\{bc\}}$ produce slightly higher values ($\beta \simeq 7.65$, $\beta_0 \simeq 57.0$). In the $\Omega_{bbc}$ sector the ordering is inverted: the equal-flavor $\Omega_{c\{bb\}}$ channel gives the largest slope ($\beta \simeq 10.1$) and intercept ($\beta_0 \simeq 109.8$), while the mixed-flavor $\Omega_{b[bc]}$ and $\Omega_{b\{bc\}}$ channels fall lower ($\beta \simeq 9.5$, $\beta_0 \simeq 106.6$). This inversion mirrors the reduced-mass inversion already noted in Sec.~\ref{subsec:mass_spectra} and underlines that, within the quark--diquark framework, the favored clustering depends on which heavy quark acts as the spectator.

In summary, the radial Regge analysis reveals a coherent picture. The $P$-wave trajectories are accurately linear, the $S$-wave trajectories exhibit a small but physically interpretable curvature at low $n_r$ that reflects the Coulomb--confinement interplay of the Cornell potential, and both slopes and intercepts scale systematically with the heavy-quark content of the baryon. The extracted parameters, $\beta \simeq 5.6$--$10.1~\mathrm{GeV}^{2}$ and $\beta_0 \simeq 55.8$--$121.1~\mathrm{GeV}^{2}$, lie within the ranges reported by earlier works, and the goodness-of-fit values ($R^{2} \simeq 0.968$--$0.999$) confirm both the internal consistency of the nonrelativistic quark--diquark framework and the robustness of the computed spectra.

\section{Conclusion}
\label{sec:conclusion}

In this work, we have carried out a systematic investigation of the mass spectra, magnetic moments, and radial Regge trajectories of the triply heavy baryons $\Omega_{ccb}$ and $\Omega_{bbc}$ within a nonrelativistic quark--diquark framework. The three-body problem is reduced to an effective two-body system for each of the three possible diquark clusterings---$b + \{cc\}$, $c + [bc]$, and $c + \{bc\}$ for $\Omega_{ccb}$, and $c + \{bb\}$, $b + [bc]$, and $b + \{bc\}$ for $\Omega_{bbc}$. The five model parameters $(m_c, m_b, \alpha_s, b, \sigma)$ have been determined by a direct fit to the measured $B_c$ meson spectrum, which reproduces both the $1S$ ground state and the $2S$ radial excitation within the experimental uncertainties. This $B_c$-anchored calibration strategy establishes a transparent link between the heavy meson and baryon sectors and eliminates the need for baryon-specific parameter tuning. Only the spin--spin component of the Breit--Fermi Hamiltonian is retained, so the spin--orbit and tensor interactions responsible for resolving the fine structure of the $L \geq 1$ multiplets, including possible $J = 5/2$ P-wave states, are not incorporated in the present framework. 

The resulting mass spectra place the $\Omega_{ccb}$ and $\Omega_{bbc}$ ground states at approximately $8.0$ and $11.0$~GeV, respectively, in agreement with previous quark-model, lattice QCD, and QCD sum rule analyses. The $1S$--$2S$ splitting is consistently close to $600$~MeV across both sectors, and the $1P$ state lies between the $1S$ and $2S$ levels as expected from a Cornell-type potential. A comparison across the three diquark clusterings reveals that the predictions spread by at most a few hundred MeV in each sector, providing a quantitative estimate of the systematic uncertainty inherent to the quark--diquark reduction. When benchmarked against the broader theoretical literature, the mixed-flavor $[bc]/\{bc\}$ diquark configurations emerge as favored for $\Omega_{ccb}$, while the equal-flavor $\{bb\}$ configuration is favored for $\Omega_{bbc}$; in both sectors, the favored clustering corresponds to a charm quark acting as spectator of a more compact heavy diquark, a pattern consistent with heavy-quark scale-separation expectations. The near-exact degeneracy of the scalar and axial-vector $bc$ diquark configurations and the strong suppression of the hyperfine splitting between the $J = 1/2$ and $J = 3/2$ baryon states are both natural consequences of the $1/\mu_{Qd}^{2}$ scaling of the chromomagnetic interaction in heavy-heavy systems.

The predicted magnetic moments, computed from the standard spin--flavor wave functions with effective quark masses that include binding corrections, yield $\mu(\Omega_{ccb}) = 0.544\,\mu_N$, $\mu(\Omega_{ccb}^{*}) = 0.718\,\mu_N$, $\mu(\Omega_{bbc}) = -0.227\,\mu_N$, and $\mu(\Omega_{bbc}^{*}) = 0.267\,\mu_N$. The sign flip between $\Omega_{bbc}$ and its spin-$3/2$ partner constitutes a distinctive experimental signature, and the global magnitude hierarchy $|\mu(\Omega_{ccb}^{*})| > |\mu(\Omega_{ccb})| > |\mu(\Omega_{bbc}^{*})| > |\mu(\Omega_{bbc})|$ is a robust consequence of the underlying heavy-quark dynamics. Comparison with seven independent calculations in the literature places our predictions comfortably within the spread of existing results in every channel.

The radial Regge analysis in the $(n_r, M^{2})$ plane exhibits two clearly distinguishable behaviors: the $P$-wave trajectories are accurately linear, with $R^{2} \geq 0.994$ in every channel, whereas the $S$-wave trajectories show a small but physically interpretable curvature at low $n_r$ ($R^{2} \simeq 0.97$--$0.99$). This $S$-wave deviation is naturally explained by the Coulombic depression of the $1S$ state within the Cornell potential and is consistent with the $M^{2} \sim n_{r}^{4/3}$ scaling reported in alternative parametrizations. The extracted slopes, $\beta \simeq 5.6$--$10.1~\mathrm{GeV}^{2}$, and intercepts, $\beta_0 \simeq 55.8$--$121.1~\mathrm{GeV}^{2}$, scale systematically with the heavy-quark content of the baryon; indeed, $\sqrt{\beta_0}$ reproduces the ratio of the physical ground-state masses at the percent level. The dependence on the diquark spin is marginal, consistent with the suppression of the chromomagnetic splitting already observed in the mass spectrum.

The present study is mainly focused on the low-lying S- and P-wave states,
which constitute the dominant low-energy excitations of triply heavy baryons within the nonrelativistic quark--diquark framework. Higher orbital excitations such as D- and F-wave states were not included, since their reliable description requires the systematic incorporation of spin--orbit and tensor interactions, which are beyond the scope of the present study. The investigation of higher-L excitations will be considered in future work. Moreover, higher orbital excitations may be more sensitive to fine-structure effects and relativistic corrections, which could require more refined treatments for a fully quantitative description of the D- and F-wave states.

Several limitations of the present approach should be acknowledged and constitute natural directions for future work. Only the spin--spin component of the Breit--Fermi Hamiltonian is retained, so the spin--orbit and tensor interactions that split $L \geq 1$ multiplets are not resolved, and the $P$-wave masses should be interpreted as spin-averaged values. Relativistic recoil corrections, including the Thomas precession, may become non-negligible for the most asymmetric clusterings ($c + \{bb\}$ and $b + \{cc\}$) and deserve a systematic semi-relativistic treatment. Production cross sections, decay widths, and lifetimes, which ultimately determine experimental detectability, have not been addressed here. Finally, the absence of experimental data on triply heavy baryons precludes a direct validation of the predictions; first-principles lattice QCD results in this sector continue to provide the most authoritative external benchmarks. The present nonrelativistic quark--diquark framework provides a comparatively simple and computationally efficient description of triply heavy baryons by reducing the three-body problem to an effective two-body system while preserving the essential features of heavy-quark dynamics. Moreover, the use of parameters constrained by the experimental $B_c$ meson spectrum reduces the model dependence associated with arbitrary parameter choices and establishes a direct connection between the heavy meson and baryon sectors. The present framework is particularly suitable for triply heavy systems, where the large heavy-quark masses suppress relativistic effects and make the nonrelativistic approximation more reliable than in light-quark baryons. The resulting predictions show overall good agreement with various quark-model, lattice QCD, and QCD sum-rule results available in the literature.

In summary, the nonrelativistic quark--diquark model constrained by the $B_c$ meson spectrum furnishes a consistent and predictive description of the triply heavy $\Omega_{ccb}$ and $\Omega_{bbc}$ baryons. The mass spectra, magnetic moments, and Regge trajectories reported here enlarge the theoretical landscape with which future measurements---particularly at LHCb---can be compared, and the identification of favored diquark clusterings offers a specific internal structure hypothesis that such measurements may in principle discriminate.

\bibliographystyle{apsrev4-2}
\bibliography{triply-heavy}

\end{document}